\documentclass[12pt,draftclsnofoot,onecolumn]{IEEEtran}
\makeatletter
\def\ps@headings{%
\def\@oddhead{\mbox{}\scriptsize\rightmark \hfil \thepage}%
\def\@evenhead{\scriptsize\thepage \hfil \leftmark\mbox{}}%
\def\@oddfoot{}%
\def\@evenfoot{}}
\makeatother
\usepackage{multirow}
\usepackage{amssymb}
\usepackage{amsmath}
\usepackage{amsfonts}
\usepackage{graphicx}
\usepackage{subfigure}
\usepackage{cite}
\usepackage{enumerate}
\usepackage{url}
\usepackage{bm}
\usepackage{clrscode}
\usepackage{subfigure}
\usepackage{amsthm}
\usepackage{mathrsfs}
\usepackage{cases}
\usepackage{CJK}
\usepackage{indentfirst}
\usepackage{algorithm}
\usepackage{algorithmic}
\allowdisplaybreaks[3]

\begin{document}

\title{Cellular UAV-to-X Communications: Design\\ and Optimization for Multi-UAV Networks}

\author{
\IEEEauthorblockN{\normalsize{Shuhang Zhang, Hongliang Zhang, {\em{Student Member}}, {\emph{IEEE}}, Boya Di, {\em{Student Member}}, {\emph{IEEE}}, and Lingyang Song, {\em{Senior Member}}, {\emph{IEEE}}}}\\
\thanks{The authors are with the School of Electronics Engineering and Computer Science, Peking University, Beijing, China. Email:$\{$shuhangzhang, hongliang.zhang, diboya, lingyang.song$\}$@pku.edu.cn.}
}
\maketitle
\setlength{\abovecaptionskip}{0pt}
\setlength{\belowcaptionskip}{-10pt}
\begin{abstract}
In this paper, we consider a single-cell cellular network with a number of cellular users~(CUs) and unmanned aerial vehicles~(UAVs), in which multiple UAVs upload their collected data to the base station (BS). Two transmission modes are considered to support the multi-UAV communications, i.e., UAV-to-infrastructure (U2I) and UAV-to-UAV~(U2U) communications. Specifically, the UAV with a high signal to noise ratio (SNR) for the U2I link uploads its collected data directly to the BS through U2I communication, while the UAV with a low SNR for the U2I link can transmit data to a nearby UAV through underlaying U2U communication for the sake of quality of service. We first propose a cooperative UAV sense-and-send protocol to enable the UAV-to-X communications, and then formulate the subchannel allocation and UAV speed optimization problem to maximize the uplink sum-rate. To solve this NP-hard problem efficiently, we decouple it into three sub-problems: U2I and cellular user~(CU) subchannel allocation, U2U subchannel allocation, and UAV speed optimization. An iterative subchannel allocation and speed optimization algorithm (ISASOA) is proposed to solve these sub-problems jointly. Simulation results show that the proposed ISASOA can upload 10\% more data than the greedy algorithm.
\end{abstract}

\newpage
\section{Introduction}
Unmanned aerial vehicle (UAV) is an emerging facility which has been effectively applied in military, public, and civil applications \cite{GJV2015}. According to BI Intelligence's report, more than 29 million UAVs are expected to be put into use in 2021~\cite{UAVmarket}. Among these applications, the use of UAV to perform sensing has been of particular interest owing to its significant advantages, such as the ability of on-demand flexible deployment, larger service coverage compared with the conventional fixed sensor nodes, and additional design degrees of freedom by exploiting its high UAV mobility~\cite{WJHRMH2017, MTA2016}. Recently, UAVs with cameras or sensors have entered the daily lives to execute various sensing tasks, e.g. air quality index monitoring~\cite{YZBSH2017}, autonomous target detection~\cite{KGM2017}, precision agriculture~\cite{AMCG2017}, and water stress quantification~\cite{SDWC2017}. The sensory data needs to be transmitted to the server for further processing, thereby posing high uplink rate requirement on the UAV communication network.

Driven by such real-time requirements, the upcoming network is committed to support UAV communication, where the collected data can be effectively transmitted~\cite{KKGM2016}. Unlike the conventional ad hoc sensor network, the sensory data can be transmitted to the 5G networks directly in a centralized way~\cite{DSLL2017}, which can greatly improve the quality of UAV communications~\cite{ACCIM2013}. In this paper, we study a single cell cellular network with a number of cellular users (CUs) and UAVs, where each UAV moves along a pre-determined trajectory to collect data, and then uploads these data to the base station (BS). However, some UAVs may located at the cell edge, and the signal to noise ration (SNR) of their communication links to the BS are low. To provide a satisfactory data rate, we enable these UAVs to transmit the sensory data to the UAVs with high SNR for the communication link to the BS as relay. The relaying UAVs save the received data in their caches and upload the data to the BS in the following time slots as described in~\cite{CMSYDH2017}. Specifically, the UAV transmissions can be supported by two basic modes, namely UAV-to-infrastructure~(U2I) and UAV-to-UAV (U2U) transmissions. The overlay U2I transmission offers direct link from UAVs with high SNR to the BS, and thus provides a high data rate~\cite{MSBD2017,ZZ2017}. In U2U transmission, a UAV with low SNR for the U2I link can set up direct communication links to the high U2I SNR UAVs bypassing the network infrastructure and share the spectrum with the U2I and CU transmissions, which provides a spectrum-efficient method to support the data relaying process~\cite{ZLS2017}.

Due to the high mobility and long transmission distance of the sensing UAVs, it is not trivial to address the following issues. Firstly, since the U2U transmissions underlay the spectrum resources of the U2I and CU transmissions, the U2I and CU transmissions may be interfered by the U2U transmissions when sharing the same subchannel. Correspondingly, the U2U transmissions are also interfered by the U2I, CU, and U2U links on the same subchannel. Moreover, different channel models are utilized for the U2I, U2U, and CU transmissions due to the different characteristics of air-to-ground, air-to-air, and ground-to-ground communications. Therefore, an \emph{efficient spectrum allocation algorithm} is required to manage the mutual interference. Secondly, to complete the data collection of the sensing tasks given time requirements, the \emph{UAV speed optimization} is necessary. Thirdly, to avoid the data loss and provide a relatively high data rate for the UAVs with low SNR for the link to the BS, an \emph{efficient communication method} is essential. In summary, the resource allocation schemes, UAV speed, and UAV transmission protocol should be properly designed to support the UAV-to-X communications.

In the literature, some works on the UAV communication network have been studied, in which UAVs work as relays or BSs. In~\cite{AEY2018}, the authors studied a 3-D UAV-BS placement to maximize the number of covered users with different quality-of-service requirements. In~\cite{MSBD2016}, the deployment of a UAV as a flying BS used to provide the fly wireless
communications was analyzed. In~\cite{ZZZ2017}, the UAV was proposed to work as a mobile BS which collected data from fixed sensor nodes on the ground. A trajectory design and power control algorithm was introduced for a UAV relay network in \cite{ZZHBS2017} to improve the reliability of transmissions. The work~\cite{MSBD2017'} investigated the scenario where UAVs served as flying BSs to provide wireless service to ground users, and optimized the downlink data rate and UAV hover duration. In~\cite{LZZ2017}, the authors proposed a hybrid network architecture with the use of UAV as a BS, which flies cyclically along the cell edge to serve the cell-edge users. Unlike most of the previous works which typically treat UAVs as relays or BSs, in our work the UAVs that relay the data from other UAVs also have their own sensing tasks, i.e. we consider the UAVs as flying mobile terminals in the UAV sensing network.

The main contributions of this paper can be summarized below.
\begin{enumerate}[(1)]
\item We construct a UAV communication network, where the UAVs can either upload the collected data via U2I communications directly or send to other UAVs by U2U communications. A cooperative UAV sense-and-send protocol is proposed to enable these communications.
\item We formulate a joint subchannel allocation and UAV speed optimization problem to maximize the uplink sum-rate of the network. We then prove that the problem is NP-hard, and decompose it into three sub-problems: U2I and CU subchannel allocation, U2U subchannel allocation, and UAV speed optimization. An efficient iterative subchannel allocation and speed optimization algorithm (ISASOA) is proposed to solve the sub-problems iteratively.
\item We compare the proposed algorithm with a greedy algorithm in simulations. The results show that the proposed ISASOA outperforms the greedy algorithm by about 10\% in terms of the uplink sum-rate.
\end{enumerate}

The rest of this paper is organized as follows. In Section~\uppercase\expandafter{\romannumeral2}, we present the system model
of the UAV sensing network. A cooperative UAV sense-and-send protocol is proposed in Section~\uppercase\expandafter{\romannumeral3} for the data collection and UAV-to-X communications. In Section~\uppercase\expandafter{\romannumeral4}, we formulate the uplink sum-rate maximization problem by optimizing the subchannel allocation and UAV speed jointly. The ISASOA is proposed in Section~\uppercase\expandafter{\romannumeral5}, followed by the corresponding analysis. Simulation results are presented in Section~\uppercase\expandafter{\romannumeral6}, and finally we conclude the paper in Section~\uppercase\expandafter{\romannumeral7}.
\vspace{-5mm}
\section{System model}
In this section, we first describe the working scenario, and then introduce the data transmission of this network. Finally, we present the channel models for U2I, U2U, and CU transmissions, respectively.\vspace{-5mm}
\subsection{Scenario Description}
\begin{figure}[!ht]
\centerline{\includegraphics[width=12cm]{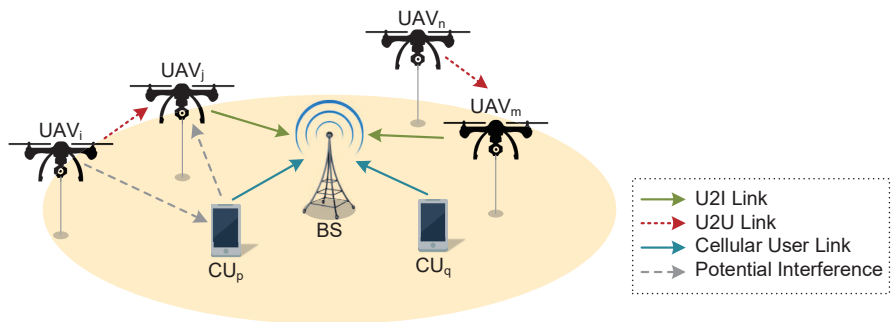}}
\caption{System model.}\vspace{-6mm}
\label{systemmodel}
\end{figure}

We consider a single cell cellular network as shown in Fig.~\ref{systemmodel}, which consists of one BS, $M$ CUs, denoted by ${\cal{M}}=\{ 1, 2, \cdots, M\}$, and $N$ UAVs, denoted by ${\cal{N}}=\{ 1, 2, \cdots, N\}$. The UAVs collect various required data with their sensors in each time slot, and the data will be sent to the BS for further processing.

We denote the location of UAV $i$ in time slot $t$ by $\textbf{l}_i(t)=(x_i(t),y_i(t),h_i(t))$, and the location of the BS by~$(0,0,H)$. Each UAV moves along a pre-determined trajectory. Let $\textbf{v}_i(t)$ be the velocity of UAV $i$ in time slot $t$. The location of UAV $i$ in time slot $t+1$ is given as $\textbf{l}_i(t+1)=\textbf{l}_i(t)+\textbf{v}_i(t)\cdot \bm{\omega}_i(t)$, where $\bm{\omega}_i(t)$ is the trajectory direction of UAV $i$ in time slot $t$. Due to the mechanical limitation, the velocity of a UAV is no more than $v_{max}$\footnote{We consider the UAVs as rotary wing UAVs which can hover in the air for some time slots. The rotary wing UAVs can move with the velocity of [0, $v_{max}$] in any time slot.}. Let $L_i$ be the length of UAV~$i$'s trajectory. With proper transmission rate requirements, the UAVs are capable to upload the sensory data to the BS with low latency. Therefore, the task completion time of a UAV can be defined as the time that it costs to complete its moving along the trajectory, which is determined by its speed in each time slot. For timely data collection, the task completion time of each UAV is required to be no more than $T$ time slots, i.e., $\sum_{t=1}^T\|\textbf{v}_i(t)\|\geq L_i, \forall i\in \cal{N}$. In time slot $t$, the distance between UAV $i$ and UAV $j$ is shown as
\begin{equation}
d_{i,j}(t)=\sqrt{\big(x_i(t)-x_j(t)\big)^2+\big(y_i(t)-y_j(t)\big)^2+\big(h_i(t)-h_j(t)\big)^2},
\end{equation}
and the distance between UAV $i$ and BS is expressed as
\begin{equation}
d_{i,BS}(t)=\sqrt{x_i(t)^2+y_i(t)^2+\big(h_i(t)-H\big)^2}.
\end{equation}
The location of CU $i$ is given as $\big(x^c_i,y^c_i,h^c_i\big)$. In this paper, we assume that the locations of the CUs are fixed in different time slots, as the mobility of the CUs are much lower than that of the UAVs. Therefore, the distance between CU $i$ and UAV $j$ can be denoted by
\begin{equation}
d_{i,j}^c(t)=\sqrt{\big(x_i^c(t)-x_j(t)\big)^2+\big(y_i^c(t)-y_j(t)\big)^2+\big(h_i^c(t)-h_j(t)\big)^2},
\end{equation}
and the distance between UAV $i$ and BS can be shown as
\begin{equation}
d_{i,BS}^c(t)=\sqrt{x_i^c(t)^2+y_i^c(t)^2+\big(h_i^c(t)-H\big)^2}.
\end{equation}
\vspace{-8mm}
\subsection{Data Transmission}
In this part, we give a brief introduction to the data transmission of this network. To provide a high data transmission rate for all the UAVs, we distinguish the UAVs with different quality of service for the link to the BS into different transmission schemes. There are two types of UAV transmission schemes in this network, namely U2I transmission and U2U transmission\footnote{We assume that the CUs only transmit data to the BS, and the device-to-device transmission between CUs is out of the scope of this network.}. A UAV may either perform U2I transmission or U2U transmission in one time slot. The criterion of adopting U2I or U2U transmission is given below.
\begin{enumerate}
\item \textbf{U2I Transmission}: A UAV with high SNR for the link to the BS performs U2I transmission in the network. It uploads its collected data to the BS directly over the assigned subchannel.
\item \textbf{U2U Transmission}: A UAV with low SNR for the link to the BS performs U2U communication to transmit the collected data to a UAV in U2I transmission scheme.
\end{enumerate}

Let $\mathcal{N}_h(t)=\{ 1, 2, \cdots, N_h(t)\}$ and $\mathcal{N}_l(t)=\{ 1, 2, \cdots, N_l(t)\}$ be the set of UAVs that perform U2I and U2U transmissions in time slot $t$, respectively, with ${\cal{N}}=\mathcal{N}_h(t) \cup \mathcal{N}_l(t)$. For the UAVs in $\mathcal{N}_h(t)$, they send the data to the BS by U2I transmissions overlaying the cellular ones. For the UAVs in $\mathcal{N}_l(t)$, the SNR of the direct communication links are low, which are difficult to provide high data rates to support timely data upload via U2I transmissions. Therefore, the UAVs send the collected data to the neighbouring UAVs with high SNR for the U2I link via U2U transmissions, which work as an underlay of the U2I and CU transmissions, and the data will be sent to the BS later by the relaying UAVs. The sensing and transmission are performed simultaneously by each UAV, and the detailed procedure will be elaborated in Section~\ref{Data Model}.

The transmission bandwidth of this network is divided into $K$ orthogonal subchannels, denoted by ${\cal{K}}=\{ 1, 2, \cdots, K\}$, and the U2U transmission UAVs work in an underlay mode, i.e., they reuse the spectrum resources with the U2I and CU transmissions. It is worthwhile to mention that a single UAV can perform U2I transmission and U2U reception over different subchannels simultaneously. For the sake of transmission quality, we assume that a subchannel can serve at most one U2I or CU link, but multiple U2U links in one time slot. In addition, to guarantee fairness among the users, we also assume that each transmission link can be allocated to no more than $\chi_{max}$ subchannels. In time slot $t$, we define a $(N_h+M)\times K$ binary U2I and CU subchannel pairing matrix $\bm{\Phi}(t)=[\phi_{i,k}(t)]$, and a $N_l\times K$ binary U2U subchannel pairing matrix $\bm{\Psi}(t)=[\psi_{i,k}(t)]$, to describe the resource allocation for CU, U2I and U2U transmissions, respectively. For $i\leq N_h$, $\phi_{i,k}(t)=1$ when subchannel $k$ is assigned to UAV $i$ for U2I transmission, otherwise $\phi_{i,k}(t)=0$. For $i> N_h$, $\phi_{i,k}(t)=1$ when subchannel $k$ is assigned to CU $i-N$ for CU transmission, otherwise $\phi_{i,k}(t)=0$.
Likewise, the value of $\psi_{i,k}(t)=1$ when subchannel $k$ is assigned to UAV $i$ for U2U transmission, otherwise $\psi_{i,k}(t)=0$.

We denote $\xi_{i,j}(t)=1$ when UAV $i$ performs U2U transmission with UAV $j$ in time slot $t$, and $\xi_{i,j}(t)=0$ otherwise. In order to avoid the high communication latency for the UAVs, the data rate of each U2U communication link should be no less $R_0$, i.e. $\sum_{k=1}^K \psi_{i,k}(t) {R_{i,j}^k(t)}\geq R_0, \forall i, j \in \mathcal{N}, \xi_{i,j}=1$.
\vspace{-5mm}
\subsection{Channel Model}\label{Channel}
In this subsection, we introduce the channel model in this network. The channel models of the U2I, CU, and U2U transmissions are different, due to the different characteristics in LoS probability and elevation angel, which will be introduced as follows, respectively.
\subsubsection{U2I Channel Model}
We use the air-to-ground propagation model which is proposed in~\cite{AKJ2014,AKL2014,AGSB2016} for the U2I transmission. In time slot $t$, the LoS and NLoS pathloss from UAV $i$ to the BS is given by
\begin{equation}\label{PL LoS}
PL_{LoS,i}(t)=L_{FS,i}(t)+20 \log(d_{i,BS}(t))+\eta_{LoS},
\end{equation}
\begin{equation}\label{PL NLoS}
PL_{NLoS,i}(t)=L_{FS,i}(t)+20 \log(d_{i,BS}(t))+ \eta_{NLoS},
\end{equation}
where $L_{FS,i}(t)$ is the free space pathloss given by $L_{FS,i}(t)=20\log(f)+20\log(\frac{4\pi}{c})$, and $f$ is the system carrier frequency. $\eta_{LoS}$ and $\eta_{NLoS}$ are additional attenuation factors due to the LoS and NLoS connections. Considering the antennas on UAVs and the BS placed vertically, the probability of LoS connection is given by\vspace{-1mm}
\begin{equation}\label{LoS Probability}\vspace{-1mm}
P_{LoS,i}(t)=\frac{1}{1+a\exp(-b(\theta_i(t)-a))},
\end{equation}
where $a$ and $b$ are constants which depend on the environment, and $\theta_i(t)=\sin^{-1}((h_i(t)-H)/d_{i,BS}(t))$ is the elevation angle. The average pathloss in dB can then be expressed as\vspace{-1mm}
\begin{equation}\label{average PL}\vspace{-1mm}
PL_{avg,i}(t)=P_{LoS,i}(t)\times PL_{Los,i}(t)+P_{NLoS,i}(t)\times PL_{NLoS,i}(t),
\end{equation}
where $P_{NLoS}(t)=1-P_{LoS}(t)$. The average received power of BS from UAV $i$ over its paired subchannel $k$ is given by\vspace{-1mm}
\begin{equation}\label{BS receive power}\vspace{-1mm}
P_{i,BS}^k(t)=\frac{P_U}{10^{PL_{avg,i}(t)/10}},
\end{equation}
where $P_U$ is the transmit power of a UAV or CU over one subchannel.
Since each subchannel can be assigned to at most one U2I or CU link, the interference to the U2I transmissions only comes from the U2U transmissions due to spectrum sharing. 
When UAV $i$ performs U2I transmission over subchannel $k$, the U2U interference is expressed as\vspace{-2mm}
\begin{equation}\label{U2I Interference}\vspace{-2mm}
I_{k,U2U}(t)=\sum_{j=1}^{N_l} \psi_{j,k}(t)P_{j,BS}^k(t).
\end{equation}
Therefore, the signal to interference plus noise ratio (SINR) of the BS over subchannel $k$ is given by\vspace{-2mm}
\begin{equation}\label{BS receive SINR}\vspace{-2mm}
\gamma_{i,BS}^{k}(t)=\frac{P_{i,BS}^k(t)}{\sigma^2+I_{k,U2U}(t)},
\end{equation}
where $\sigma^2$ is the variance of additive white Gaussian noise (AWGN) with zero mean. The data rate that BS receives from UAV~$i$ over subchannel $k$ is shown as\vspace{-2mm}
\begin{equation}\label{BS receive rate}\vspace{-2mm}
R_{i,BS}^{k}(t)=\log_2(1+\gamma_{i,BS}^{k}(t)).
\end{equation}
\subsubsection{CU Channel Model}
We utilize the macrocell pathloss model as proposed in~\cite{3GPP25.996}. For CU~$i$, the pathloss in dB can be expressed by
\begin{equation}\label{CU pathloss}\vspace{-2mm}
PL_{i,C}^k(t) = -55.9 +38 \log(d_{i,BS}^c(t))+(24.5+1.5f/925)\log(f).
\end{equation}
When CU $i$ transmits signals to BS, the received power is expressed as\vspace{-2mm}
\begin{equation}\label{CU Received_power}\vspace{-2mm}
P_{i,C}^k(t) = \frac{P_U}{10^{PL_{i,C}^k(t)/10}}.
\end{equation}
We denote the set of UAVs that share subchannel $k$ with CU $i$ by $U_{i}=\{m|\psi_{m,k}(t)=1, \forall m\in \mathcal{N}_e\}$, and the received power at the BS over subchannel $k$ is shown as\vspace{-2mm}
\begin{equation}\label{CU Received_signal}\vspace{-2mm}
y_{i,j}^k(t)=\sqrt{P_{i,C}^k(t)} + \sum\limits_{m\in{U_{i}}}\sqrt{P_{m,BS}^k(t)} + n_{j}^k(t),
\end{equation}
where $n_{j}^k(t)$ is the AWGN with zero mean and $\sigma^2$ variance. Therefore, the received signal at the BS over subchannel $k$ can be given by\vspace{-2mm}
\begin{equation}\label{BS receive CU SINR}\vspace{-2mm}
\gamma_{i,BS}^{k}(t)=\frac{P_{i,C}^k(t)}{\sigma^2+I_{k,U2U}(t)},
\end{equation}
where $I_{k,U2U}(t)=\sum_{j=1}^N \psi_{j,k}(t)P_{j,BS}^k(t)$ is the U2U interference. The data rate for CU $i$ over subchannel $k$ is expressed as\vspace{-2mm}
\begin{equation}\label{BS receive CU rate}\vspace{-2mm}
R_{i,BS}^{k}(t)=\log_2(1+\gamma_{i,BS}^{k}(t)).
\end{equation}
\subsubsection{U2U Channel Model}
For U2U communication, free-space channel model is utilized. When UAV $i$ transmits signals to UAV $j$ over subchannel $k$, the received power at UAV $j$ from UAV $i$ is expressed as\vspace{-2mm}
\begin{equation}\label{UAV Received_power}\vspace{-2mm}
P_{i,j}^k(t) = P_U G (d_{i,j}(t))^{-\alpha},
\end{equation}
where $G$ is the constant power gains factor introduced by amplifier and antenna, and $(d_{i,j}(t))^{-\alpha}$ is the pathloss.
Define the set of UAVs and CUs that share subchannel $k$ with UAV $i$ as $W_{i}=\{m|\psi_{m,k}(t)=1, \forall m\in \mathcal{N}_e\setminus i\}\cup \{m|\phi_{m,k}(t)=1\}$. The received signal at UAV $j$ over subchannel $k$ is then given by\vspace{-2mm}
\begin{equation}\label{UAV Received_signal}\vspace{-2mm}
y_{i,j}^k(t)=\sqrt{P_{i,j}^k(t)}+ \sum\limits_{m\in{W_{i}}}\sqrt{P_{m,j}^k(t)}+ n_{j}^k(t),
\end{equation}
where $P_{m}^k(t)$ is the received power at UAV $j$ from the UAVs and CUs in $W_{i}$, and $n_{j}^k(t)$ is the AWGN with zero mean and $\sigma^2$ variance.
The interference from UAV $m$ to UAV $j$ over subchannel $k$ is shown as\vspace{-2mm}
\begin{equation}\label{UAV U2U Interference}\vspace{-2mm}
I_{m,UAV}^k(t)=(\phi_{m,k}(t)+\psi_{m,k}(t))P_U (d_{m,j}(t))^{-\alpha}.
\end{equation}
According to the channel reciprocity, the interference from CU $m$ to UAV $j$ over subchannel $k$ can be expressed as\vspace{-2mm}
\begin{equation}\label{CU U2U Interference}\vspace{-2mm}
I_{m,C}^k(t)=\phi_{m,k}(t)\frac{P_U}{10^{PL_{avg,j}^m(t)/10}},
\end{equation}
where $PL_{avg,j}^m(t)$ is the average pathloss from UAV $j$ to CU $m$, which can be derived from equation (\ref{PL LoS})-(\ref{average PL}).
The SINR at UAV $j$ over subchannel $k$ is shown as
\begin{equation}\label{UAV Received_SINR}
\gamma_{i,j}^k(t)=\frac{P_U (d_{i,j}(t))^{-\alpha} }{\sigma^2+\sum\limits_{m=1, m\neq i}^{N_l+N_h} I_{m,UAV}^k(t)+\sum\limits_{m=1}^{M}I_{m,C}^k(t)}.
\end{equation}
When UAV $i$ transmits its data to UAV $j$ over subchannel $k$ via U2U transmission, the data rate is given by\vspace{-2mm}
\begin{equation}\label{UAV Received_rate}\vspace{-2mm}
R_{i,j}^k(t)=\log_2(1+\gamma_{i,j}^k(t)).
\end{equation}

\section{Cooperative UAV Sense-and-Send Protocol}\label{Data Model}
\begin{figure}[!ht]
\centerline{\includegraphics[width=12cm]{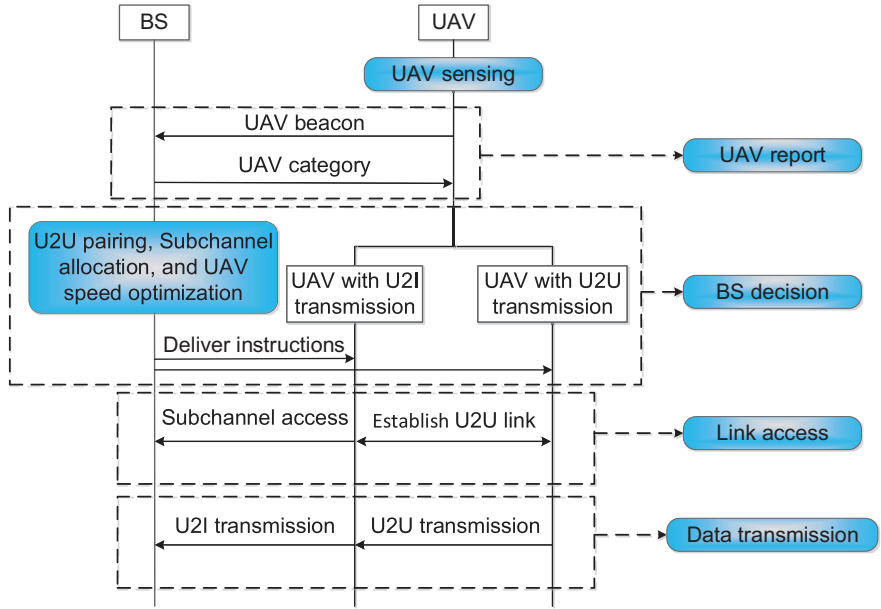}}
\caption{Cooperative UAV sense-and-send protocol.}
\label{protocol}
\end{figure}

In this section, we propose a cooperative UAV sense-and-send protocol that supports the UAV data collections and UAV-to-X transmissions in this network. As illustrated in Fig.~\ref{protocol}, in each time slot, the UAVs first collect the sensory data of their tasks. They then send beacons to the BS over the control channel, and the BS categorizes the UAVs according to the received SNR. Afterwards, the BS performs U2U pairing, subchannel allocation and UAV speed optimization for the UAVs in the network, and sends the results to the UAVs. After receiving the results, the UAVs establish the transmission links, and perform U2I and U2U transmissions according to the arrangement of the BS. To better describe the protocol, we divide each time slot into five steps: UAV sensing, UAV report, BS decision, link access, and data transmission, and introduce them in details in the following.

\textbf{UAV Sensing:} In the UAV sensing step, the UAVs perform data sensing and save the collected data in their caches. The communication module is turned off in the UAV sensing step.

\textbf{UAV Report:} After the UAV sensing step, the UAVs stop data collection and send beacons to the BS. The beacon of each UAV contains its ID and current location, and is sent to the BS over the control channel in a time-division manner. When receiving the beacons of the UAVs, the BS categorizes the UAVs into U2I and U2U transmissions according to the received SNR. The UAVs with high SNR are considered to perform U2I transmission, and the UAVs low SNR are considered to perform U2U transmission.

\textbf{BS Decision:} After categorizing the UAVs, the BS pairs the UAVs that perform U2U transmission with their closest UAV with U2I transmission. The BS then performs subchannel allocation and UAV speed optimization with our proposed algorithm described in Section~\uppercase\expandafter{\romannumeral5}. Afterwards, the results are sent to the UAVs over the control channel.

\textbf{Link Access:} When the control signals from the BS are sent to the UAVs, the UAVs start to move with the optimized speed and transmit over the allocated subchannel. The UAVs with U2I transmissions access the allocated subchannels provided by the BS, and the UAVs with U2U transmissions establish the U2U links with the corresponding UAV relays over the allocated subchannels.

\textbf{Data Transmission:} The UAVs start to transmit data to the corresponding target after the communication links are established successfully. The data transmission step lasts until the end of the time slot.
\vspace{-5mm}
\section{Problem Formulation}
In this section, we first formulate the joint subchannel allocation and UAV speed optimization problem, and prove that the optimization problem is NP-hard, which can not be solved directly within polynomial time. Therefore, in the next part, we decouple it into three sub-problems, and elaborate them separately.\vspace{-3mm}
\subsection{Joint Subchannel Allocation and UAV Speed Optimization Problem Formulation}
Since all the data collected by the UAVs needs to be sent to the BS, the uplink sum-rate of this network is one key index to evaluate the performance of this network. We aim to maximize the uplink sum-rate by optimizing the subchannel allocation and UAV speed variables $\bm{\omega}(t)$, $\bm{\Phi}(t)$, and $\|\bm{v}_i(t)\|$. The joint subchannel allocation and UAV speed optimization problem can be formulated as follows:
\begin{subequations} \label{system_optimization}
\begin{align}
\mathop{\max }\limits_{{\{\|\textbf{v}_i(t)\|\},\{\textbf{$\bm{\Phi}(t)$}\},\{\textbf{$\bm{\Psi}(t)$}\}}}&\sum_{k=1}^{K}\sum_{i=1}^{N_h+M} \phi_{i,k}(t) R_{i,BS}^{k}(t),\label{system_1}\\
\textbf{\emph{s.t. }}&\sum_{k=1}^K \psi_{i,k}(t) {R_{i,j}^k(t)}\geq R_0, \forall i, j \in \mathcal{N}, \xi_{i,j}=1,\label{system_3}\\
&\|\bm{v}_i(t)\|\leq v_{max}, \forall i\in \mathcal{N},\label{system_4}\\
&\sum_{t=1}^T\|\textbf{v}_i(t)\|\geq L_i, \forall i\in \mathcal{N},\label{system_10}\\
&\sum_{i=1}^{N_h+M} \phi_{i,k}(t)\leq 1, \forall k \in \mathcal{K},\label{system_6}\\
&\sum_{k=1}^K \phi_{i,k}(t)\leq \chi_{max}, \forall i \in \mathcal{N}_h(t) \cup \mathcal{M},\label{system_7}\\
&\sum_{k=1}^K \psi_{i,k}(t)\leq \chi_{max}, \forall i \in \mathcal{N}_l(t),\label{system_8}\\
&\phi_{i,k}(t)\in\{0, 1\}, \psi_{i,k}(t)\in\{0, 1\}, \forall i \in \mathcal{N}\cup \mathcal{M}, k \in \mathcal{K}.\label{system_9}
\end{align}
\end{subequations}
The minimum U2U transmission rate satisfies constraint (\ref{system_3}). (\ref{system_4}) is the maximum speed constraint for the UAVs, and~(\ref{system_10}) shows that the task completion time of each UAV is no more than $T$ time slots. Constraint~(\ref{system_6}) implies that each subchannel can be allocated to at most one UAV that performs U2I transmission or CU. Each UAV and CU can be paired with at most $\chi_{max}$ subchannels, which is given in constraints~(\ref{system_7}) and~(\ref{system_8}). In the following theorem, we will prove that optimization problem (\ref{system_optimization}) is NP-hard.

\textbf{Theorem 1:} Problem (\ref{system_optimization}) is NP-hard.
\begin{proof}
See Appendix A.
\end{proof}
\subsection{Problem Decomposition}\label{sub-problems}
Since problem (17) is NP-hard, to tackle this problem efficiently, we decouple problem~$(\ref{system_optimization})$ into three sub-problems, i.e., U2I and CU subchannel allocation, U2U subchannel allocation, and UAV speed optimization sub-problems. In the U2I and CU subchannel allocation sub-problem, the U2U subchannel matching matrix $\bm{\Psi}(t)$ and the UAV speed $\{\|\bm{v}_i(t)\|\}$ are considered to be fixed. Therefore, the U2I and CU subchannel allocation sub-problem is written as
\begin{subequations} \label{U2I subchannel}
\begin{align}
\mathop{\max }\limits_{{\textbf{$\bm{\Phi}(t)$}}}&\sum_{k=1}^{K}\sum_{i=1}^{N_h+M} \phi_{i,k}(t) R_{i,BS}^{k}(t),\label{U2I1}\\
\textbf{\emph{s.t. }}&\sum_{i=1}^{N_h+M} \phi_{i,k}(t)\leq 1, \forall k \in \mathcal{K},\\
&\sum_{k=1}^K \phi_{i,k}(t)\leq \chi_{max}, \forall i \in \mathcal{N}_h(t) \cup \mathcal{M},\\
&\phi_{i,k}(t)\in\{0, 1\}, \forall i \in \mathcal{N}_h(t) \cup \mathcal{M}, k \in \cal{K}.\label{U2I14}
\end{align}
\end{subequations}
Given the U2I and CU subchannel pairing matrix $\bm{\Phi}(t)$ and the UAV speed $\{\|\bm{v}_i(t)\|\}$, the U2U subchannel allocation sub-problem can be written as
\begin{subequations} \label{U2U subchannel}
\begin{align}
\mathop{\max }\limits_{{\textbf{$\bm{\Psi}(t)$}}}&\sum_{k=1}^{K}\sum_{i=1}^{N_h+M} \phi_{i,k}(t) R_{i,BS}^{k}(t),\label{U2U1}\\
\textbf{\emph{s.t. }}&\sum_{k=1}^K \psi_{i,k}(t) {R_{i,j}^k(t)}\geq R_0, \forall i, j \in \mathcal{N}_l(t), \xi_{i,j}=1,\label{U2U2}\\
&\sum_{k=1}^K \psi_{i,k}(t)\leq \chi_{max}, \forall i \in \mathcal{N}_l(t),\label{U2U3}\\
&\psi_{i,k}(t)\in\{0, 1\}, \forall i \in \mathcal{N}_l(t), k\in \mathcal{K}\label{U2U4}.
\end{align}
\end{subequations}
Similarly, when the subchannel pairing matrices $\bm{\Phi}(t)$ and $\bm{\Psi}(t)$ are given, the UAV speed optimization sub-problem can be expressed by
\begin{subequations}\label{Velocity Sub-problems}
\begin{align}
\mathop{\max }\limits_{{\{\|\textbf{v}_i(t)\|\}}}&\sum_{k=1}^{K}\sum_{i=1}^{N_h+M} \phi_{i,k}(t) R_{i,BS}^{k}(t),\label{Trajectory Sub-problem}\\
&\sum_{k=1}^K \psi_{i,k}(t) {R_{i,j}^k(t)}\geq R_0, \forall i, j \in \mathcal{N}, \xi_{i,j}=1,\label{velocity_0}\\
&\|\bm{v}_i(t)\|\leq v_{max}, \forall i\in \mathcal{N},\label{velocity_1}\\
&\sum_{t=1}^T\|\textbf{v}_i(t)\|\geq L_i, \forall i\in \mathcal{N}.\label{velocity_2}
\end{align}
\end{subequations}
\section{Joint Subchannel Allocation and UAV Speed Optimization}
In this section, we propose an effective method i.e., ISASOA to obtain a sub-optimal solution of problem (\ref{system_optimization}) by solving its three sub-problems (\ref{U2I subchannel}), (\ref{U2U subchannel}), and (\ref{Velocity Sub-problems}) iteratively. The U2I and CU subchannel allocation sub-problem (\ref{U2I subchannel}) can be relaxed to a standard linear programming problem, which can be solved by existing convex techniques, for example, CVX. We then utilize the branch-and-bound method to solve the non-convex U2U subchannel allocation sub-problem~(\ref{U2U subchannel}). For the UAV speed optimization sub-problem (\ref{Velocity Sub-problems}), we discuss the feasible region and convert it into a convex problem, which can be solved by existing convex techniques. Iterations of solving the three sub-problems are performed until the objective function converges to a constant. In the following, we first elaborate on the algorithms of solving the three sub-problems respectively. Afterwards, we will provide the ISASOA, and discuss its convergence and complexity.
\subsection{U2I and CU Subchannel Allocation Algorithm}\label{U2I subsection}
In this subsection, we give a detailed description of the U2I and CU subchannel allocation algorithm. As shown in Section \ref{sub-problems}, the decoupled sub-problem (\ref{U2I subchannel}) is an integer programming problem. To make the problem more tractable, we relax the variables $\bm{\Phi}(t)$ into continuous values, and the relaxed problem is expressed as
\begin{subequations} \label{U2I subchannel2}
\begin{align}
\mathop{\max }\limits_{{\textbf{$\bm{\Phi}(t)$}}}&\sum_{k=1}^{K}\sum_{i=1}^{N_h+M} \phi_{i,k}(t) R_{i,BS}^{k}(t),\label{U2I12}\\
\textbf{\emph{s.t. }}&\sum_{i=1}^{N_h+M} \phi_{i,k}(t)\leq 1, \forall k \in \mathcal{K},\label{U2I2}\\
&\sum_{k=1}^K \phi_{i,k}(t)\leq \chi_{max}, \forall i \in \mathcal{N}_h(t) \cup \mathcal{M},\label{U2I3}\\
&0\leq \phi_{i,k}(t)\leq 1, \forall i \in \mathcal{N}_h(t) \cup \mathcal{M}, k\in \mathcal{K}.\label{U2I4}
\end{align}
\end{subequations}

When we substitute (\ref{BS receive SINR}), (\ref{BS receive rate}), (\ref{BS receive CU SINR}), and (\ref{BS receive CU rate}) into (\ref{U2I12}), it can be observed that the pairing matrix $\bm{\Phi}(t)$ is not relevant with $R_{i,BS}^{k}(t)$. Therefore, $R_{i,BS}^{k}(t)$ is fixed in this sub-problem. Note that function (\ref{U2I12}) is linear with respect to the optimization variables $\bm{\Phi}(t)$, and equation~(\ref{U2I2}),~(\ref{U2I3}), and~(\ref{U2I4}) are all linear. Thus, problem (\ref{U2I subchannel2}) is a standard linear programming problem, which can be solved efficiently by utilizing the existing optimization techniques such as CVX~\cite{CVX}. In what follows, we will prove that the solution of the relaxed problem (\ref{U2I subchannel2}) is also the one of the original problem (\ref{U2I subchannel}).

\textbf{Theorem 2:} All the variables in $\bm{\Phi}(t)$ are met with 0 or 1 for the solution of problem (\ref{U2I subchannel2}).
\begin{proof}
See Appendix B.
\end{proof}
As shown in Theorem 2, the solution of problem (\ref{U2I subchannel2}) is either 0 or 1, which satisfies the constraint (\ref{U2I14}) of the original problem. Therefore, the relaxation of variable $\phi_{i,k}(t)$ does not affect the solution of the sub-problem (\ref{U2I subchannel}). The solution of the relaxed problem (\ref{U2I subchannel2}) with CVX method is equivalent to the solution of problem (\ref{U2I subchannel}).
\vspace{-3mm}
\subsection{U2U Subchannel Allocation Algorithm}\label{U2U subsection}
In this subsection, we focus on solving the U2U subchannel allocation sub-problem (\ref{U2U subchannel}).
We first substitute (\ref{U2I Interference}), (\ref{BS receive SINR}), (\ref{BS receive rate}), (\ref{BS receive CU SINR}), and (\ref{BS receive CU rate}) into (\ref{U2U1}), and the objective function is given by
\begin{equation}\label{Expanded Objective}
\begin{split}
\mathop{\max }\limits_{{\textbf{$\bm{\Psi}(t)$}}}&\sum_{k=1}^{K}\Bigg(\sum_{i=1}^{N_h} \phi_{i,k}(t) \log_2\left(1+\frac{P_{i,BS}^k(t)}{\sigma^2+\sum_{j=1}^{N_l} \psi_{j,k}(t)P_{j,BS}^k(t)}\right)+\\&\sum_{i=N_h+1}^{N_h+M} \phi_{i,k}(t) \log_2\left(1+\frac{P_{i,C}^k(t)}{\sigma^2+\sum_{j=1}^{N_l} \psi_{j,k}(t)P_{j,BS}^k(t)}\right)\Bigg).
\end{split}
\end{equation}
When substituting (\ref{UAV Received_SINR}) and (\ref{UAV Received_rate}) into constraint~(\ref{U2U2}), it can be expanded as
\begin{equation}\label{Expanded U2U Restriction}
\begin{split}
R_{U2U,i}=\sum_{k=1}^K \psi_{i,k}(t) \log_2\left(1+\frac{P_U (d_{i,j}(t))^{-\alpha} \|g_{i,k}\|^2}{A+\sum\limits_{m=1, m\neq i}^{N_l} \psi_{m,k}(t)P_U (d_{m,j}(t))^{-\alpha}\|g_{m,k}\|^2}\right) \geq\frac{F_i}{t_0}, \\ \forall i, j \in N, \xi_{i,j}=1,
\end{split}
\end{equation}
where $A=\sigma^2+\sum\limits_{n=1}^{N_h} \phi_{n,k}(t)I_{m,UAV}^k(t)+\sum\limits_{m=N_h+1}^{N_h+M} \phi_{m,k}(t)I_{m,C}^k(t)$ is fixed in this sub-problem. Problem (\ref{U2U subchannel}) is a \emph{0-1} programming problem, which has been proved to be NP-hard \cite{P1976}. In addition, due to the interference from different U2U links, the continuity relaxed problem of~(\ref{U2U subchannel}) is still non-convex with respect to $\bm{\Psi}(t)$. Therefore, problem (\ref{U2U subchannel}) cannot be solved by the existing convex techniques. To solve problem (\ref{U2U subchannel}) efficiently, we utilize the branch-and-bound method~\cite{LS2006}.

To facilitate understanding of the branch-and-bound algorithm, we first introduce important concepts of fixed and unfixed variables.

\textbf{Definition 1:} When the value of a variable that corresponds to the optimal solution is ensured, we define it as a \textbf{fixed variable}. Otherwise, it is an \textbf{unfixed variable}.

The solution space of U2U subchannel pairing matrix $\bm{\Psi}(t)$ can be considered as a binary tree. Each node of the binary tree contains the information of all the variables in $\bm{\Psi}(t)$. At the root node, all the variables in $\bm{\Psi}(t)$ are unfixed. The value of an unfixed variable at a father node can be either 0 or 1, which branches the node into two child nodes. Our objective is to search the binary tree for the optimal solution of problem (\ref{U2U subchannel}). The key idea of the branch-and-bound method is to prune the infeasible branches and approach the optimal solution efficiently.

At the beginning of the algorithm, we obtain a feasible solution of problem~(\ref{U2U subchannel}) by a proposed low-complexity feasible solution searching (LFSS) method, and set it as the lower bound of the solution. We then start to search the optimal solution of problem~(\ref{U2U subchannel}) in the binary tree from its root node. On each node, the branch-and-bound method consists two steps: bound calculation and variable fixation. In the bound calculation step, we evaluate the upper bound of the objective function and the bounds of the constraints separately to prune the branches that can not achieve a feasible solution above the lower bound of the solution. In the variable fixation step, we fix the variables which has only one feasible value that satisfies the bound requirements in the bound calculation step. We then search the node that contains the newly fixed variables, and continue the two steps of bound calculation and variable fixation. The algorithm terminates when we obtain a node with all the variables fixed. In what follows, we first introduce the LFSS method to achieve the initial feasible solution, and then describe the bound calculation and variable fixation process at each node in detail. Finally, we summarize the branch-and-bound method.

\subsubsection{Initial Feasible Solution Search}
In what follows, we propose the LFSS method to obtain a feasible solution of problem (\ref{U2U subchannel}) efficiently. Each UAV that performs U2U transmission requests a subchannel until its minimum U2U rate threshold is satisfied, and the BS assigns the requested subchannel to the corresponding UAV in the LFSS. The detailed description is shown in Algorithm \ref{Feasible_Alg}.

Given the U2I and CU subchannel assignment, each UAV that performs U2U transmission can make a list of data rate that it may achieve from every subchannel without considering the potential U2U interference. The UAVs then sort the subchannels in descending order of achievable rate. We then calculate the data rate of each U2U link with U2I, CU, and U2U interference when the UAVs are assigned to their most preferred subchannels. If the data rate of an UAV is still below the minimum threshold, the UAV will be assigned to its most preferred subchannel which has not been paired with. The subchannel assignment ends when the minimum U2U rate threshold (\ref{U2U3}) is satisfied by every UAV that performs U2U transmission. Finally, we adopt the current U2U subchannel pairing result as the initial feasible solution.
\begin{algorithm}[htb]
\caption{Initial Feasible Solution for U2U Subchannel Allocation.}
\begin{algorithmic}[1]\label{Feasible_Alg}
\STATE {Each UAV that performs U2U transmission calculates its data rate over every subchannel with U2I and CU interference;}
\STATE {Each UAV sorts the subchannels in descending order of achievable rate;}
\STATE {Assign the UAVs with their most preferred subchannel;}
\STATE {Calculate the data rate of each U2U link with U2I, CU, and U2U interference;}
\STATE {\textbf{While} The data rate of an UAV does not satisfy U2U rate constraint (\ref{U2U3})}
\STATE {\quad Assign the UAV to its most preferred subchannel that has not been paired;}
\STATE {\textbf{End While}}
\STATE {Set the current U2U-subchannel pairing result as the initial feasible solution;}
\end{algorithmic}
\end{algorithm}
\subsubsection{Bound Calculation}
In this part, we describe the process of bound calculation at each node. After the initialization step, we start bound calculation from the root node, in which all the variables in $\bm{\Psi}(t)$ are unfixed, i.e., the value of each $\psi_{i,k}(t)$ in the optimal solution is unknown. We first define a branch pruning operation which is performed in the following bound calculation step.

\textbf{Definition 2:} When a node is \textbf{fathomed}, all its child nodes can not be the optimal solution of the problem.

We calculate the bounds of the objective function and the constraints separately. For simplicity, we denote the objective function with U2U subchannel matrix by $f(\bm{\Psi}(t))$, and the lower bound of the solution by $f^{lb}$. In what follows, we will elaborate the detailed steps of the bound calculation at each node.

\textbf{Step 1 Objective Bound Calculation:} The upper bound of the objective function (\ref{U2U1}) is given as
\begin{equation}
\begin{split}
\bar{f}=&\sum_{k=1}^{K}\Bigg(\sum_{i=1}^{N_h} \phi_{i,k}(t) \log_2\left(1+\frac{P_{i,BS}^k(t)}{\sigma^2+\sum_{j=1}^{N_l} \psi_{j,k}^F(t)P_{j,BS}^k(t)}\right)+\\&\sum_{i=N_h+1}^{N_h+M} \phi_{i,k}(t) \log_2\left(1+\frac{P_{i,C}^k(t)}{\sigma^2+\sum_{j=1}^{N_l} \psi_{j,k}^F(t)P_{j,BS}^k(t)}\right)\Bigg),
\end{split}
\end{equation}
where $\psi_{j,k}^F(t)$ is the fixed variables in the current node, i.e., we ignore the U2U interference of the unfixed variables. If the upper bound of the current node is below the lower bound of the solution, i.e., $\bar{f}<f^{lb}$, we fathom the current node and backtrack to an unfathomed node with unfixed variable. If the current node is not fathomed by the objective function bound calculation, we move to step 2 to check the bounds of the constraints.

\textbf{Step 2 Constraint Bounds Calculation:} For each UAV that performs U2U transmission, the upper bound of its U2U rate needs to be larger than the minimum U2U rate threshold. The upper bound of U2U rate for UAV $i$ is achieved when we set all the unfixed variables of UAV $i$ as 1, and all the unfixed variables of other UAVs as 0, which can be expressed as
\begin{equation}
\begin{split}
\bar{R}_{U2U,i}=\sum_{k=1}^K \psi_{i,k}(t)|_{\{\psi_{i,k}^U(t)=1\}} \log_2\left(1+\frac{P_U (d_{i,j}(t))^{-\alpha}\|g_{i,k}\|^2}{A+\sum\limits_{m=1, m\neq i}^{N_l} \psi_{m,k}^F(t)P_U (d_{m,j}(t))^{-\alpha}\|g_{m,k}\|^2}\right), \\ \forall i, j \in N, \xi_{i,j}=1,
\end{split}
\end{equation}
where $\psi_{m,k}^U(t)$ is the unfixed variables in the current node. If $\exists \xi_{i,j}=1, \bar{R}_{U2U,i}<\frac{F_i}{t_0}$, the minimum U2U rate threshold can not be satisfied, and the current node is fathomed. Moreover, if there exists a UAV $i$ that does not satisfy constraint (\ref{U2U3}), i.e., $\sum_{k=1}^K \psi_{i,k}(t)> \chi_{max}, \forall i \in N$, the current node is also fathomed. We then backtrack to an unfathomed node in the binary tree and perform bound calculation at the new node.

In the bound calculation procedure, if the objective function of a U2U subchannel pairing matrix $f(\tilde{\bm{\Psi}}(t))$ is found to be larger than the lower bound of the solution $f^{lb}$, and $\tilde{\bm{\Psi}}(t)$ satisfies all the constraints, we replace the lower bound of the solution with $f^{lb}=f(\tilde{\bm{\Psi}}(t))$ to improve the algorithm efficiency. A higher lower bound of the solution helps us to prune the infeasible branches more efficiently.
\subsubsection{Variable Fixation}
For a node that is not fathomed in the bound calculation steps, we try to prune the branches by fixing the unfixed variables as follows. The variable fixation is completed in two steps, namely objective fixation and U2U constraint fixation.

\textbf{Step 1 Objective Fixation:} In the objective fixation process, we denote the reduction of the upper bound when fixing a free variable $\psi_{i,k}(t)$ at 0 or 1 by $p_{i,k}^0$ and $p_{i,k}^1$, respectively. For each unfixed variable $\psi_{i,k}(t)$, we compute $p_{i,k}^0$ and $p_{i,k}^1$ associated with the upper bound $\bar{f}$. If $\bar{f}-p_{i,k}^0\leq f_{opt}$, it means that when we set $\psi_{i,k}(t)= 0$, the upper bound of the child node will fall below the temporary feasible solution. Therefore, we prune the branch of $\psi_{i,k}(t)= 0$, and fix $\psi_{i,k}(t)= 1$. Similarly, if $\bar{f}-p_{i,k}^1\leq f_{opt}$, we prune the branch of $\psi_{i,k}(t)= 1$, and fix $\psi_{i,k}(t)= 0$.

\textbf{Step 2 U2U Constraint Fixation:} In the U2U constraint fixation process, we denote the U2U rate upper bound reduction for UAV $i$ when fixing a free variable $\psi_{i,k}(t)$ at 0 by $q_{i,k}^0$. If inequality $\bar{R}_{U2U,i}-q_{i,k}^0<\frac{F_i}{t_0}$ is satisfied, it means that only when subchannel $k$ is assigned to UAV $i$, the minimum U2U rate threshold of UAV $i$ is possible to be satisfied. Therefore, we prune the branch of $\psi_{i,k}(t)=0$ and fix $\psi_{i,k}(t)=1$.

In the objective fixation step and the U2U constraint fixation step, variable $\psi_{i,k}(t)$ may be fixed at different values, which implies that neither of the two child nodes satisfy the objective bound relation and the constraint bound relation simultaneously. Therefore, we fathom the current node and backtrack to an unfathomed node with unfixed variable.

After performing the variable fixation step of the current node, if at least one unfixed variable is fixed at a certain value in the above procedure, we move to the corresponding child node, and continue the algorithm by performing bound calculation and variable fixation at the new node. Otherwise, we generate two new nodes by setting an unfixed variable at $\psi_{i,k}(t)=0$ and $\psi_{i,k}(t)=1$, respectively. We then move to one of the two nodes and continue the algorithm. The branch-and-bound algorithm is accomplished when all variables have been fixed, and the fixed variables are the final solution.

The branch-and-bound method that solves the U2U subchannel allocation sub-problem (\ref{U2U subchannel}) is summarized as Algorithm \ref{Alg1}.

\begin{algorithm}[!thp]
\caption{Branch-and-Bound Method for U2U Subchannel Allocation.}
\begin{algorithmic}[1]\label{Alg1}
\REQUIRE ~~\\ 
The U2I subchannel allocation matrix $\bm{\Phi}(t)$; The UAV trajectories $\bm{\omega}(t)$;
\ENSURE ~~\\ 
The U2U subchannel allocation matrix $\bm{\Psi}(t)$;
\STATE {\textbf{Initialization:} Compute an initial feasible solution $\bm{\Psi}(t)$ to problem (\ref{U2U subchannel}) and set it as the lower bound of the solution;}
\STATE {Perform bound calculation and variable fixation at the root node;}
\STATE {\textbf{While} Not all variables have been fixed}
\STATE {\quad Bound calculation;}
\STATE {\quad\quad\textbf{If} The bound constraints can not be satisfied}
\STATE {\quad\quad\quad Fathom the current node and backtrack to an unfathomed node with unfixed variable;}
\STATE {\quad\quad \textbf{End If}}
\STATE {\quad Variable fixation;}
\STATE {\quad\quad\textbf{If} At least one variable can be fixed}
\STATE {\quad\quad\quad \textbf{Go to} the node with newly fixed variable;}
\STATE {\quad\quad\textbf{Else} Generate two new nodes by setting an unfixed variable $\psi_{i,k}(t)=0$ and $\psi_{i,k}(t)=1$;} \STATE {\quad\quad \textbf{Go to} one of the two nodes firstly;}
\STATE {\quad\quad \textbf{End If}}
\STATE {\textbf{End While}}
\STATE {The fixed variables are the final output of $\bm{\Psi}(t)$;}
\end{algorithmic}
\end{algorithm}

\subsection{UAV Speed Optimization Algorithm}\label{Trajectory subsection}
In the following, we will introduce how to solve the UAV speed optimization sub-problem~(\ref{Velocity Sub-problems}). Note that in problem~(\ref{Velocity Sub-problems}), the speed optimization of a pair of U2U transmitting and receiving UAVs are related with constraint~(\ref{velocity_0}), but the speed optimization of different UAVs that perform U2I transmission are independent. Therefore, the speed optimization of the UAVs can be separated into two types: non-U2U participated UAVs and U2U participated UAVs that contains the transmitting UAVs and the corresponding receiving UAVs.
\subsubsection{Non-U2U Participated UAV Speed Optimization}\label{Non-U2U Section}
For non-U2U participated UAVs, constraint~(\ref{velocity_0}) is not considered. We denote the length of trajectory that UAV $i$ has moved before time slot $t$ by $\mathcal{L}_i(t)$. To satisfy constraint~(\ref{velocity_1}) and (\ref{velocity_2}), the length of trajectory that UAV $i$ needs to move along in the following time slots should be no more than the number of following time slots $T-t-1$ times the maximum UAV speed $v_{max}$, i.e. $L_i-\mathcal{L}_i(t+1)<v_{max}\times (T-t-1)$. Therefore, the feasible range of UAV $i$'s speed in time slot $t$ is $\min\{0,L_i-\mathcal{L}_i(t)-v_{max}\times (T-t-1)\}\leq \|\bm{v}_i(t)\| \leq v_{max}$. Problem~(\ref{Velocity Sub-problems}) can be simplified as
\begin{subequations}\label{Non-U2U Velocity}
\begin{align}
\mathop{\max }\limits_{{\|\textbf{v}_i(t)\|}}&\sum_{k=1}^{K}\sum_{i=1}^{N_h+M} \phi_{i,k}(t) R_{i,BS}^{k}(t),\\
&\min\{0,L_i-\mathcal{L}_i(t)-v_{max}\times (T-t-1)\}\leq \|\bm{v}_i(t)\| \leq v_{max}.
\end{align}
\end{subequations}
Note that the moving distance of a UAV in one time slot is much shorter than the length of its trajectory, i.e. $v_{max}\ll L_i$. We assume that the LoS and NLoS pathloss does not change prominently in the single time slot, and the uplink rate is determined by the the probability of the LoS and NLoS connections given in (\ref{LoS Probability}), which is a convex function. Therefore, problem~(\ref{Non-U2U Velocity}) is approximated as a convex problem, and can be solved with existed convex optimization methods.
\subsubsection{U2U Participated UAV Speed Optimization}
In this part, we introduce the speed optimization of a pair of UAVs: UAV $i$ and UAV $j$, with $\xi_{i,j}=1$. In time slot $t$, UAV $i$ performs U2U transmission and send the collected data to UAV $j$. UAV $j$ receives the data from UAV $i$, and performs U2I transmission simultaneously. As described in Section~\ref{Non-U2U Section}, constraint~(\ref{velocity_1}) and~(\ref{velocity_2}) can be simplified as $\min\{0,L_i-\mathcal{L}_i(t)-v_{max}\times (T-t-1)\}\leq \|\bm{v}_i(t)\| \leq v_{max}$, and $\min\{0,L_j-\mathcal{L}_j(t)-v_{max}\times (T-t-1)\}\leq \|\bm{v}_j(t)\| \leq v_{max}$ for UAV $i$ and UAV $j$, respectively. Given the the subchannel pairing matrices $\bm{\Phi}(t)$ and $\bm{\Psi}(t)$, the U2U rate constraint~(\ref{velocity_0}) can be transformed to a distance constraint. When substituting~(\ref{UAV Received_SINR}) and~(\ref{UAV Received_rate}) into~(\ref{velocity_0}), the U2U rate constraint can be shown as
\begin{subequations}\label{Distance constaint on U2U rate}
\begin{align}
d_{i,j}(t)\leq \frac{P_U \|g_{i,k}\|^2}{\left(\sigma^2+\sum\limits_{m=1, m\neq i}^{N_l+N_h} I_{m,UAV}^k(t)+\sum\limits_{m=1}^{M} I_{m,C}^k(t)\right)\left(2^{\frac{F_i}{t_0\times \sum_{k=1}^K \psi_{i,k}(t)}}-1\right)}.
\end{align}
\end{subequations}

Since the U2U transmission distance is much larger than the moving distance of a UAV in one time slot, i.e., $d_{i,j}(t)\gg v_{max}$. The interference can be approximated to a constant. Therefore, the right side of equation (\ref{Distance constaint on U2U rate}) can be regarded as a constant, denoted by $d_{i,j}^{max}$ for simplicity. Given the feasible speed range of UAV $i$ and the maximum distance between UAV $i$ and UAV~$j$, i.e. $d_{i,j}^{max}$, a feasible speed range of UAV $j$ in time slot $t$ can be obtained, which is written as $v_j(t)^{min}\leq \|\bm{v}_j(t)\| \leq v_j(t)^{max}$. The UAV speed optimization sub-problem is reformulated as
\begin{subequations}\label{U2U Velocity}
\begin{align}
\mathop{\max }\limits_{{\|\textbf{v}_i(t)\|}}&\sum_{k=1}^{K}\sum_{i=1}^{N_h+M} \phi_{i,k}(t) R_{i,BS}^{k}(t),\\
&\min\{0,L_i-\mathcal{L}_i(t)-v_{max}\times (T-t-1)\}\leq \|\bm{v}_i(t)\| \leq v_{max},\\
&\min\{0,L_j-\mathcal{L}_j(t)-v_{max}\times (T-t-1)\}\leq \|\bm{v}_j(t)\| \leq v_{max},\\
&v_j(t)^{min}\leq \|\bm{v}_j(t)\| \leq v_j(t)^{max}.
\end{align}
\end{subequations}
Similar with problem (\ref{Non-U2U Velocity}), problem (\ref{U2U Velocity}) can also be considered as a convex problem, which can be solved with existed convex optimization methods.
\subsection{Iterative Subchannel Allocation and UAV Speed Optimization Algorithm}\label{Iterative Algorithm Subsection}
In this subsection, we introduce the ISASOA to solve problem (\ref{system_optimization}), where U2I and CU subchannel allocation, U2U subchannel allocation, and UAV speed optimization sub-problems are solved iteratively. In time slot $t$, we denote the optimization objective function after the $r$th iteration by $\mathcal{R}\Big(\bm{\Phi}^r(t), \bm{\Psi}^r(t), \bm{v}^r(t)\Big)$. In iteration $r$, the U2I and CU subchannel allocation matrix $\bm{\Phi}(t)$, the U2U subchannel allocation matrix $\bm{\Psi}(t)$, and the UAV speed variable of UAV $i$ are denoted by $\bm{\Phi}^r(t)$, $\bm{\Psi}^r(t)$, and $v_i^r(t)$, respectively. The process of the iterative algorithm for each single time slot is summarized in Algorithm \ref{Ite_Alg}.

\begin{algorithm}[htb]
\caption{Iterative Subchannel Allocation and UAV Speed Optimization Algorithm.}
\begin{algorithmic}[1]\label{Ite_Alg}
\STATE {\textbf{Initialization:} Set $r=0$, $\bm{\Phi}^0(t)=\{0\}$, $\bm{\Psi}^0(t)=\{0\}$, $\omega_i^0(t)=\{0\}, \forall i \in I(t)$;}
\STATE {\textbf{While} $\mathcal{R}\Big(\bm{\Phi}^r(t), \bm{\Psi}^r(t), \bm{\omega}^r(t)\Big)-\mathcal{R}\Big(\bm{\Phi}^{r-1}(t), \bm{\Psi}^{r-1}(t), \bm{\omega}^{r-1}(t)\Big)>\epsilon$}
\STATE {\quad $r=r+1$;}
\STATE {\quad Solve U2I and CU subchannel allocation sub-problem (\ref{U2I subchannel}), given $\bm{\Psi}^{r-1}(t)$ and $\bm{v}^{r-1}(t)$;}
\STATE {\quad Solve U2U subchannel allocation sub-problem (\ref{U2U subchannel}), given $\bm{\Phi}^{r}(t)$ and $\bm{v}^{r-1}(t)$;}
\STATE {\quad Solve UAV speed optimization sub-problem (\ref{Velocity Sub-problems}), given $\bm{\Phi}^{r}(t)$ and $\bm{\Psi}^{r}(t)$;}
\STATE {\textbf{End While}}
\STATE {\textbf{Output:}$\bm{\Phi}^r(t), \bm{\Psi}^r(t), \bm{v}^r(t)$.}
\end{algorithmic}
\end{algorithm}

In time slot $t$, we firstly set the initial condition, where all the subchannels are vacant, and the speed of all the UAVs are given as a fixed value $v_0$, i.e. $\bm{\Phi}^0(t)=\{0\}$, $\bm{\Psi}^0(t)=\{0\}$, and $v_i^0(t)=\{v_0\}, \forall i \in \mathcal{N}$. We then perform iterations of subchannel allocation and UAV speed optimization until the objective function converges. In each iteration, the U2I and CU subchannel allocation is performed first with the U2U subchannel pairing and UAV speed results given in the last iteration, and the U2I and CU subchannel pairing variables are updated. Next, the U2U subchannel allocation is performed as shown in Section \ref{U2U subsection}, with the UAV speed obtained in the last iteration and the U2I and CU subchannel pairing results. Afterwards, we perform UAV speed optimization as described in Section \ref{Trajectory subsection}, given the subchannel pairing results. When an iteration is completed, we will compare the values of the objective function obtained in the last two iterations. If the difference between the values is less than a pre-set error tolerant threshold~$\epsilon$, the algorithm terminates and the results of subchannel pairing and UAV speed optimization are obtained. Otherwise, the ISASOA will continue.

In the following, we will discuss the convergency and complexity of the proposed ISASOA.

\textbf{Theorem 3:} The proposed ISASOA is convergent.
\begin{proof}
In the $(r+1)$th iteration, we first perform U2I and CU subchannel allocation, and the optimal U2I and CU subchannel allocation solution is obtained with the given $\bm{\Psi}^r(t)$ and $v_i^r(t)$. Therefore, we have\vspace{-3mm}
\begin{equation}
\mathcal{R}\Big(\bm{\Phi}^{r+1}(t), \bm{\Psi}^r(t), \bm{v}^r(t)\Big)\geq\mathcal{R}\Big(\bm{\Phi}^r(t), \bm{\Psi}^r(t), \bm{v}^r(t)\Big),
\end{equation}
i.e., the total rate of U2I and CU transmissions does not decrease with the U2I and CU subchannel allocation in the $(r+1)$th iteration.
When solving U2U subchannel allocation, we give the optimal solution of $\bm{\Psi}^{r+1}(t)$ with $\bm{\Phi}^{r+1}(t)$ and $\bm{v}^r(t)$. The relation between $\mathcal{R}\Big(\bm{\Phi}^{r+1}(t), \bm{\Psi}^{r+1}(t), \bm{v}^r(t)\Big)$ and $\mathcal{R}\Big(\bm{\Phi}^{r+1}(t), \bm{\Psi}^r(t), \bm{v}^r(t)\Big)$ can then be expressed as
\begin{equation}
\begin{split}
\mathcal{R}\Big(\bm{\Phi}^{r+1}(t), \bm{\Psi}^{r+1}(t), \bm{v}^r(t)\Big){\geq}\mathcal{R}\Big(\bm{\Phi}^{r+1}(t), \bm{\Psi}^r(t), \bm{v}^r(t)\Big).
\end{split}
\end{equation}
The optimal speed for the UAVs with $\bm{\Phi}^r(t)$ and $\bm{\Psi}^r(t)$ are obtained in the UAV speed optimization algorithm, which can be expressed as
\begin{equation}
\mathcal{R}\Big(\bm{\Phi}^{r+1}(t), \bm{\Psi}^{r+1}(t), \bm{v}^{r+1}(t)\Big)\geq\mathcal{R}\Big(\bm{\Phi}^{r+1}(t), \bm{\Psi}^{r+1}(t), \bm{v}^r(t)\Big).
\end{equation}

In the $r+1$th iteration, we have the following inequation
\begin{equation}\label{Iteration Inequation}
\begin{split}
&\mathcal{R}\Big(\bm{\Phi}^{r+1}(t), \bm{\Psi}^{r+1}(t), \bm{v}^{r+1}(t)\Big)\geq\mathcal{R}\Big(\bm{\Phi}^{r+1}(t), \bm{\Psi}^{r+1}(t), \bm{v}^r(t)\Big)\geq\\
&\mathcal{R}\Big(\bm{\Phi}^{r+1}(t), \bm{\Psi}^r(t), \bm{v}^r(t)\Big)\geq\mathcal{R}\Big(\bm{\Phi}^r(t), \bm{\Psi}^r(t), \bm{v}^r(t)\Big).
\end{split}
\end{equation}
As shown in (\ref{Iteration Inequation}), the objective function does not decrease in each iteration. It is known that such a network has a capacity bound, and the uplink sum-rate can not increase unlimitedly. Therefore, the objective function has an upper bound, and will converge to a constant after limited iterations, i.e. the proposed ISASOA is convergent.
\end{proof}

\textbf{Theorem 4:} The complexity of the proposed ISASOA is $O((N_h(t)+M)\times2^{N_l(t)})$.
\begin{proof}
The complexity of the proposed ISASOA is the number of iterations times the complexity of iteration. As shown in Algorithm \ref{Ite_Alg}, the objective function increases for at least $\epsilon$ in each iteration. We denote the average uplink sum-rate of the initial solution by $\bar{R}_0(N_h(t),M)$, and the average uplink sum-rate of the ISASOA by $\bar{R}(N_h(t),M)$. The number of iteration is no more than $(\bar{R}(N_h(t),M)-\bar{R}_0(N_h(t),M))/\epsilon$. The increment of the uplink sum-rate can be expressed as $(\bar{R}(N_h(t),M)-\bar{R}_0(N_h(t),M))=(N_h(t)+M)\log_2\left(\frac{1+\bar{\gamma}_{I}}{1+\bar{\gamma}_{0}}\right)$, where $\bar{\gamma}_{I}$ is the average SNR of UAVs that perform U2I transmission and CUs with ISASOA, and $\bar{\gamma}_{0}$ is the average SNR of UAVs that perform U2I transmission and CUs with the initial solution. Therefore, the number of iterations is given as $C\times(N_h(t)+M)$, where $C$ is a constant.

In each iteration, the U2I subchannel allocation is solved directly with convex problem solutions. The U2U subchannel allocation is solved with branch-and-bound method, with the complexity being $O(2^{N_l(t)})$. The speed of different UAVs are optimized with convex optimization methods, with a complexity of $O(N_h(t)+N_l(t))$. Therefore, the complexity of each iteration is $O(2^{N_l(t)})$, and the complexity of the proposed ISASOA is $O((N_h(t)+M)\times2^{N_l(t)})$.
\end{proof}
\vspace{-6mm}
\section{Simulation Results}
In this section, we evaluate the performance of the proposed ISASOA. The selection of the simulation parameters are based on the existing works and 3GPP specifications\cite{3GPPR12,ZZ2017}. In this simulation, the location of the UAVs are randomly and uniformly distributed in an 3-dimension area of 2 km $\times$ 2 km $\times$ $h_{max}$, where $h_{max}$ is the maximum possible height for the UAVs. To study the impact of UAV height on the performance of this network, we simulate two scenarios with $h_{max}$ being 100 m and 200 m, respectively. The direction of the pre-determined trajectory for each UAV is given randomly. A SNR threshold $\gamma_{th}$ is given to distinguish the UAVs that perform U2I and U2U transmissions. The UAVs with the SNR for U2I links being larger than $\gamma_{th}$ are considered to perform U2I transmission, and the UAVs with the SNR for the U2I links being lower than $\gamma_{th}$ are considered to perform U2U transmission. All curves are generated with over 1000 instances of the proposed algorithm. The simulation parameters are listed in Table \uppercase\expandafter{\romannumeral2}. We compare the proposed algorithm with a greedy subchannel allocation algorithm as proposed in~\cite{ZDSL2017}. In the greedy algorithm scheme, the subchannel allocation is performed based on matching theory, and the UAV speed is the same as the proposed ISASOA scheme. The maximum possible height for the UAVs in the greedy algorithm is set as 200 m.

\begin{table}[!t]
\centering
\caption{Simulation Parameters}
\begin{tabular}{|c|c|}
\hline
\textbf{Parameter} & \textbf{Value}\\
\hline
Number of subchannels $K$ & 10\\
\hline
Number of UAVs that perform U2U transmission $N_l$ & 5 \\
\hline
Number of UAVs $N$ & 20 \\
\hline
Number of CUs $M$ & 5\\
\hline
Transmission power $P_U$ & 23 dBm\\
\hline
Noise variance $\sigma^2$ & -96 dBm\\
\hline
Center frequency & 1 GHz\\
\hline
Power gains factor $G$ & -31.5 dB\\
\hline
$X_{max}$ & 2\\
\hline
Algorithm convergence threshold $\epsilon$ & 0.1\\
\hline
U2I channel parameter $\eta_{LoS}$ & 1\\
\hline
U2I channel parameter $\eta_{NLoS}$ & 20\\
\hline
U2I channel parameter $a$ &12\\
\hline
U2I channel parameter $b$ &0.135\\
\hline
U2U pathloss coefficient $\alpha$ &2\\
\hline
Maximum UAV speed $v_{max}$ &10 m/time slot\\
\hline
Length of trajectory $L_i$ &300 m\\
\hline
Minimum U2U rate $R_0$ &10 bit/(s$\times$Hz)\\
\hline
SNR threshold $\gamma_{th}$ & 10 dB\\
\hline
\end{tabular}
\end{table}

\begin{figure}[!h]
\centerline{\includegraphics[width=12cm]{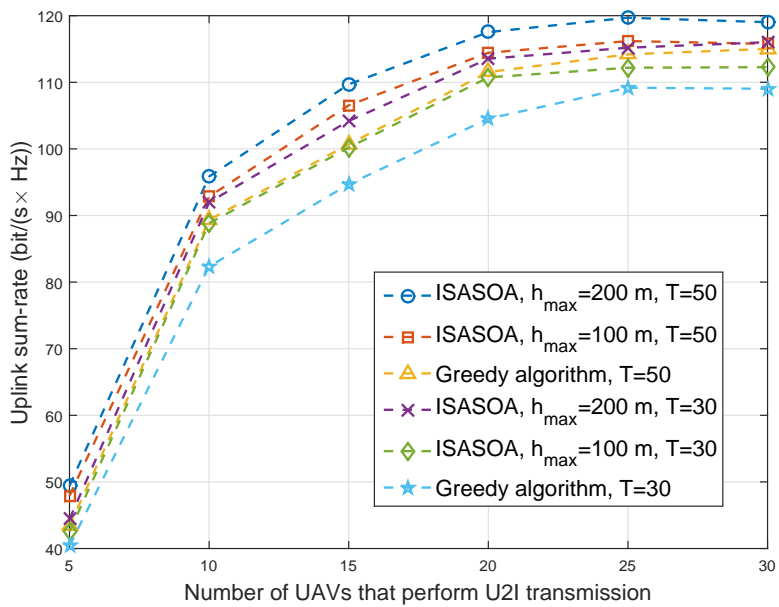}}
\vspace{-3mm}
\caption{Number of UAVs that perform U2I transmission vs. Uplink sum-rate.}\label{Fig2}
\vspace{-8mm}
\end{figure}
Fig. \ref{Fig2} depicts the uplink sum-rate with different number of UAVs that perform U2I transmission. In the proposed ISASOA scheme, the difference between $T=50$ and $T=30$ in terms of the uplink sum-rate is about 7\%. It is shown that a larger task completion time $T$ corresponds to a higher uplink sum-rate, because the UAVs have larger degree of freedom on the optimization of their speeds with a looser time constraint. The scenario with $h_{max}=200$ m has about 3\% higher uplink sum-rate than the scenario with $h_{max}=100$ m. The performance gap between the two scenarios is mainly affected by the U2I pathloss caused by different LoS and NLoS probabilities. The uplink sum-rate with the ISASOA is 10\% larger than that of the greedy algorithm on average, due to the efficient U2I and U2U subchannel allocation. All the six curves show that the uplink sum-rate of U2I and CU transmissions increases with the number of UAVs, and the growth becomes slower as $N$ increases due to the saturation of network capacity.

\begin{figure}[!t]
\centerline{\includegraphics[width=12cm]{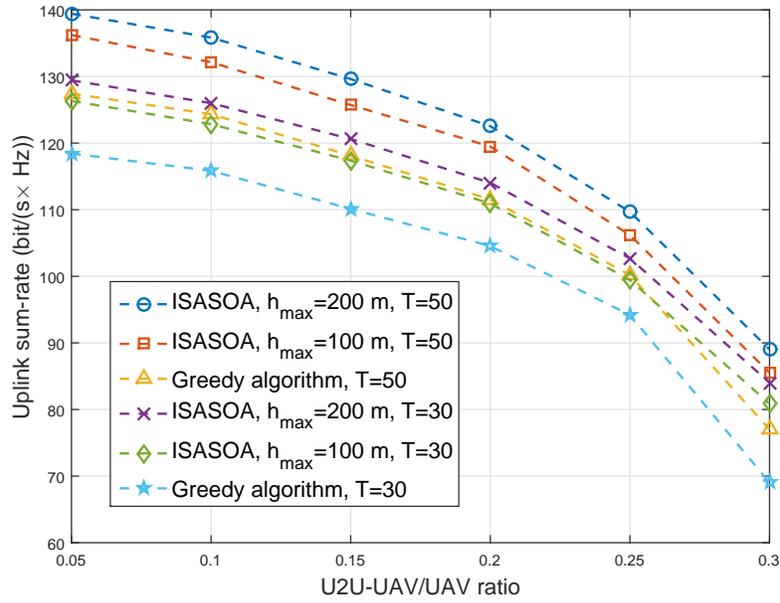}}
\vspace{-3mm}
\caption{U2U-UAV/UAV ratio vs. Uplink sum-rate.}\label{Fig7}
\vspace{-8mm}
\end{figure}
Fig. \ref{Fig7} shows the uplink sum-rate with different U2U-UAV/UAV ratio, when the number of UAVs is set as 20. It is shown that the uplink sum-rate decreases with more UAVs that perform U2U transmission in the network, and the descent rate is larger with more UAVs that perform U2U transmission. A larger U2U-UAV/UAV ratio not only reduces the number of UAVs that perform U2I transmission, but also leads to a larger number of U2U receiving UAVs. Therefore, more UAVs that perform U2I transmission are restricted by the U2U transmission rate constraint, and cannot move with the speed that corresponds to the maximum rate for the U2I links.

\begin{figure}[!t]
\centerline{\includegraphics[width=12cm]{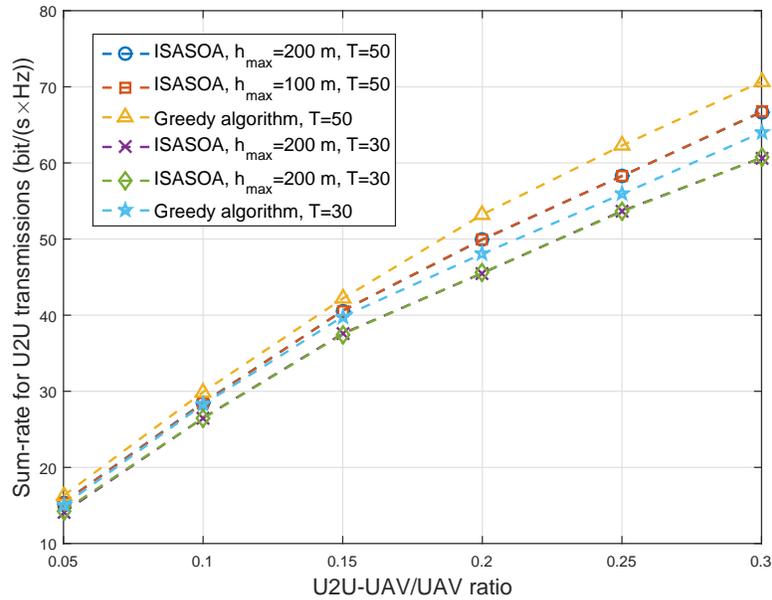}}
\vspace{-3mm}
\caption{U2U-UAV/UAV ratio vs. Sum-rate for U2U transmissions.}\label{Fig3}
\vspace{-8mm}
\end{figure}
Fig. \ref{Fig3} illustrates the relation between the U2U-UAV/UAV ratio and the sum-rate for U2U transmissions, with the number of UAVs set at 20. The total U2U transmission rate increases with a larger U2U-UAV/UAV ratio, but the rate of the increment decreases with a larger U2U-UAV/UAV ratio, i.e. the average U2U transmission rate decreases with more UAVs that perform U2I transmission in the network. The reason is that with the increment of UAVs that perform U2I transmission, the U2U-to-U2U interference raises rapidly, which reduces the data rate for a U2U link. There is no significant difference between the ISASOA scheme with different $h_{max}$ in terms of the total U2U transmission rate, since the U2U transmission rate for each link is only determined by the distance between the U2U transmitting and receiving UAVs. Note that the average U2U transmission rate is always above the U2U rate threshold within the simulation range. For the greedy algorithm scheme, the total U2U transmission rate is 5\% higher than the ISASOA scheme, but a higher U2U transmission rate squeezes the network capacity for the U2I transmissions.

\begin{figure}[!t]
\centerline{\includegraphics[width=12cm]{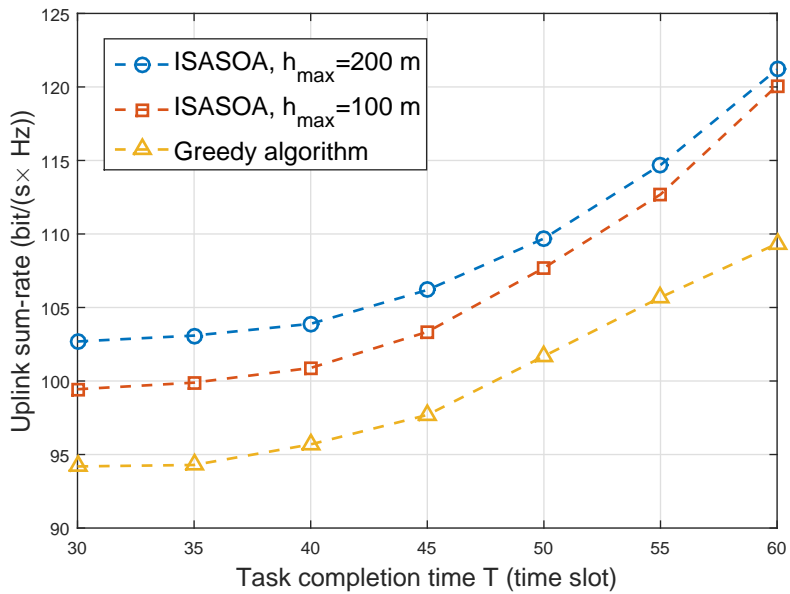}}
\vspace{-3mm}
\caption{Minimum task completion time $T$ vs. Uplink sum-rate.}\label{Fig8}
\vspace{-8mm}
\end{figure}
In Fig. \ref{Fig8}, we give the relation between the task completion time $T$ and the uplink sum-rate. The uplink sum-rate increases with a larger minimum task completion time $T$, and the rate of change increases with $T$. The scheme with $h_{max}=200$ m has a larger uplink sum-rate than the scheme with $h_{max}=100$ m due to a higher probability of LoS U2I transmission. The performance gap decreases when $T$ becomes larger, because the UAVs with $h_{max}=100$ can stay for a longer time at the locations with relatively high LoS transmission probability. The greedy algorithm is about 10\% lower than the ISASOA scheme. It can be referred that the uplink sum-rate is affected by the delay tolerance of the data collection.

\begin{figure}[!t]
\centerline{\includegraphics[width=12cm]{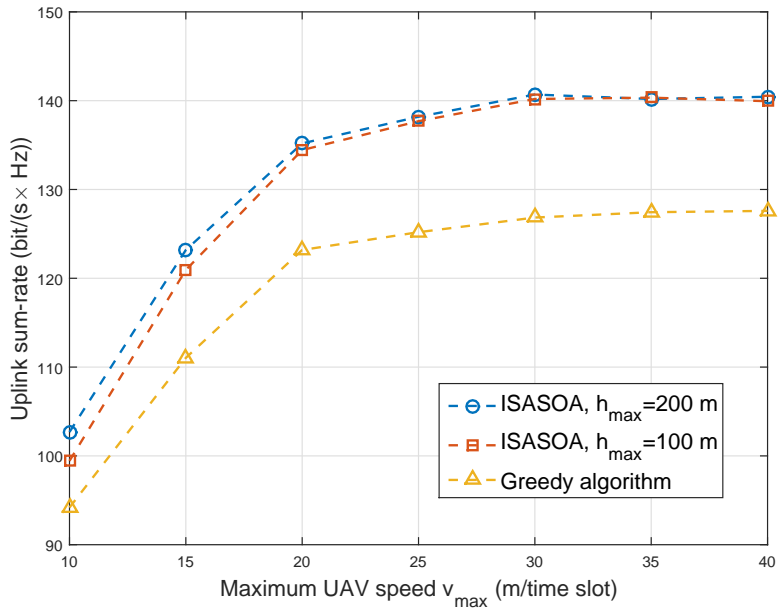}}
\vspace{-3mm}
\caption{Maximum UAV speed $v_{max}$ vs. Uplink sum-rate.}\label{Fig9}
\vspace{-8mm}
\end{figure}
In Fig. \ref{Fig9}, the uplink sum-rate is shown with different maximum UAV speed $v_{max}$. A larger maximum UAV speed provides the UAVs a larger degree of freedom on the UAV speed optimization. It is shown that the uplink sum-rate increases significantly with the maximum UAV speed when $v_{max}\leq 20$ m/time slot. The uplink sum-rate turns stable when $v_{max}> 30$ m/time slot, because speed is not the main restriction on the uplink sum-rate when the maximum UAV speed is sufficiently large. The difference between the $h_{max}=200$ and $h_{max}=100$ schemes decreases with the increment of $v_{max}$, and the greedy algorithm is about 10\% lower than the ISASOA scheme within the simulation range.
\section{Conclusion}
In this paper, we studied a single cell multi-UAV network, where multiple UAVs upload their collected data to the BS via U2I and U2U transmissions. We proposed a cooperative UAV sense-and-send protocol and formulated a joint subchannel allocation and UAV speed optimization problem to improve the uplink sum-rate of the network. To solve the NP-hard problem, we decoupled it into three sub-problems: U2I and CU subchannel allocation, U2U subchannel allocation, and UAV speed optimization. The three sub-problems were then solved with optimization methods, and the novel ISASOA was proposed to obtain a convergent solution of this problem. This network can be extended to a multiple cell scenario with BS association and inter-cell interference consideration. Simulation results showed that the uplink sum-rate decreases with a tighter task completion time constraint, and the proposed ISASOA can achieve about 10\% more uplink sum-rate than the greedy algorithm.
\vspace{-3mm}
\begin{appendices}
\section{Proof of Theorem 1}
\begin{proof}
In this appendix, we proof that problem $(\ref{system_optimization})$ is NP-hard even when we do not perform UAV speed optimization. We construct an instance of problem $(\ref{system_optimization})$ where each subchannel can only serve no more than one U2U link and one U2I or CU link simultaneously. Let $\mathcal{N}_c$, $\mathcal{N}_e$, and $\cal{K}$ be three disjoint sets of UAVs that perform U2I transmission and CU, UAVs that perform U2U transmission, and subchannels, respectively, with $|\mathcal{N}_c|=N_h$, $|\mathcal{N}_e|=N_l$, and $|\cal{K}|=\text{K}$. Set $\mathcal{N}_c$, $\mathcal{N}_e$, and $\cal{K}$ satisfy $\mathcal{N}_c \cap \mathcal{N}_e=\varnothing, \mathcal{N}_c \cap \cal{K}=\varnothing$, and $\mathcal{N}_e \cap \cal{K}=\varnothing$. Let $\cal{P}$ be a collection of ordered triples $\cal{P} \subseteq \mathcal{N}_c\times\mathcal{N}_e\times\cal{K}$, where each element in $\cal{P}$ consists a CU/UAV that perform U2I transmission, a UAV that perform U2U transmission, and a subchannel, i.e., $P_i$ = $\left(N_{c,i}, N_{e,i}, K_{i}\right)\in \cal{P}$. To be convenient, we set $L=\min\{N_h, N_l, K\}$. There exists $\cal{P}' \subseteq \cal{P}$ that holds: $(1) \left|\cal{P}'\right|$ = $\text{L}$; $(2)$ for any two distinct triples $\left(N_{c,i}, N_{e,i}, K_{i}\right)\in \cal{P}'$ and $\left(N_{c,j}, N_{e,j}, K_{j}\right)\in \cal{P}'$, we have $i \neq j$. Therefore, $\cal{P}'$ is a three dimension matching (3-DM). Since 3-DM problem has been proved to be NP-complete in \cite{GJ1979}, the constructed instance of problem is also NP-complete. Thus, the problem in $(\ref{system_optimization})$ is NP-hard~\cite{M2012}.
\end{proof}\vspace{-6mm}
\section{Proof of Theorem 2}
\begin{proof}
We assume that the solution of (\ref{U2I subchannel2}) contains a variable $\phi_{i,k}(t)$ with $0<\phi_{i,k}(t)<1$. For simplicity, we denote the slope of $\phi_{i,k}(t)$ in the objective function (\ref{U2I12}) by $X_{i,k}=\log_2\left(1+\frac{P_{i,BS}^k(t)}{\sigma^2+I_{k,U2U}(t)}\right)$, where $X_{i,k}>0, \forall i\in N, k\in K$. When the objective function is maximized, at least one of the constraints between (\ref{U2I2}) and (\ref{U2I3}) is met with equality. In the following, we separate the problem into two conditions, and discuss them successively.\vspace{-5mm}
\subsection{Only One Constraint is Met with Equality}
Without loss of generality, we assume that only (\ref{U2I2}) is met with equality. Since $\phi_{i,k}(t)$ is not an integer, there exists another variable $\phi_{j,k}(t)$ that is also non-integer to meet the constraint equality of (\ref{U2I2}). We assume that $X_{i,k}>X_{j,k}$. When we increase $\phi_{i,k}(t)$ and decrease $\phi_{j,k}(t)$ within the constraint, the objective function will be improved. Thus, the solution with $0<\phi_{i,k}(t)<1$ is not the optimal solution.\vspace{-5mm}
\subsection{Both (\ref{U2I2}) and (\ref{U2I3}) are Met With Equality}
When both (\ref{U2I2}) and (\ref{U2I3}) are met with equality, there are at least three more variables that are non-integer to meet the constraint equality. We denote the other three variables by $\phi_{j,k}(t)$, $\phi_{i,m}(t)$, and $\phi_{j,m}(t)$. If $X_{i,k}+X_{j,m}>X_{i,m}+X_{j,k}$, when we increase $\phi_{i,k}(t)$ and $\phi_{j,m}(t)$, and decrease $\phi_{j,k}(t)$ and $\phi_{i,m}(t)$, the objective function will be improved. If $X_{i,k}+X_{j,m}<X_{i,m}+X_{j,k}$, the opponent adjustment will improve the objective function. As a result, the current solution is not the optimal one.

In conclusion, the solution that contains $0<\phi_{i,k}(t)<1$ is not the optimal one. When the optimal solution of (\ref{U2I subchannel2}) is achieved, all the variables in $\bm{\Phi}(t)$ are either 0 or 1.
\end{proof}
\end{appendices}

\end{document}